\documentclass[10pt]{article}
\usepackage{latexsym,amsmath,amssymb,amsbsy,amstext,amscd,amsfonts}
\usepackage{color}
\usepackage{graphics,graphicx}
 \usepackage{latexsym,amsmath,amssymb,amsbsy,amstext,amscd,amsfonts}
 \usepackage{graphics,graphicx}
 \usepackage{color}
 \usepackage{tabularx}
 \usepackage{multirow}
\usepackage{booktabs}

\input{tcilatex}

\begin{document}

\date{\today}
\title{On modeling hydraulic fracture in proper variables: \\
stiffness, accuracy, sensitivity }
\author{Gennady Mishuris$^{(1)}$, Michal Wrobel$^{(1,2)}$,  Alexander
Linkov$^{(2,3)}$  \\[4mm]
\textit{$^{(1)}$ Institute of Mathematical and Physical Sciences,
Aberystwyth University, } \\
\textit{Ceredigion SY23 3BZ, Wales U.K.,} \\
\textit{$^{(2)}$ Eurotech Sp. z o.o., } \\
\textit{ul. Wojska Polskiego 3, 39-300 Mielec, Poland,} \\
\textit{$^{(3)}$ Department of Mathematics}\\
{\it Rzeszow University of Technology, al. Powsta\'nc\'ow Warszawy 12,}\\ {\it 35 - 959, Rzeszow, Poland.
 }}
\maketitle

\begin{abstract}
The problem of hydraulic fracture propagation is considered by using its
recently suggested modified formulation in terms of the particle velocity,
the opening in the proper degree, appropriate spatial coordinates and
$\varepsilon$-regularization. We show that the formulation may serve for significant
increasing the efficiency of numerical tracing the fracture propagation. Its
advantages are illustrated by re-visiting the Nordgren problem. It is shown
that the modified formulation facilitates (i) possibility to have various
stiffness of differential equations resulting after spatial discretization,
(ii) obtaining highly accurate and stable numerical results with moderate
computational effort, and (iii) sensitivity analysis. The exposition is
extensively illustrated by numerical examples.
\end{abstract}

\section{Introduction}

Hydraulic fracturing is a widely used method serving to increase the
linear size of an area of fluid or gas flow (see, e.g. the reviews
in papers \cite {Howard and Fast (1969)}, \cite{Economides2000},
\cite{Adachi-et-Al-2007}). In view of practical significance of the
method, numerous papers have been published on the theory and
numerical modeling of
hydraulic fractures starting from the first publications \cite%
{Khristianovich and Zheltov (1955)}, \cite{Carter}, \cite{Perkins}, \cite%
{Geertsma and de Klerk (1969)}, \cite{Howard and Fast (1969)}, \cite%
{Nordgren}, \cite{Spence&Sharp} and \cite{Nolte (1988)}. The general
formulation of the problem is well established (e.g.
\cite{Adachi-et-Al-2007}). It includes (i) the fluid equations for
flow of incompressible viscous fluid in the narrow channel; (ii) the
solid mechanics (commonly static linear elasticity) equations defining the
dependence of the channel's height on the pressure acting on the
walls of the channel; (iii) equations of fracture mechanics defining
the possibility of the fracture propagation and the trajectory of
the fracture contour. Additional equations for proppant movement are
added when accounting for the proppant injected at some stage of
fracturing.

The formulated mathematical problem is difficult from the
computational point of view because of three major complicating
factors: strong nonlinearity even in the simplest case of Newtonian
fluid, caused by the fact that the channel's height (fracture
opening) raised to some power enters the Poiseuille equation as a
multiplier by the unknown flux; moving boundaries of the fluid front
and fracture contour; and, in general, the need to check the
fracture conditions for finding the value and direction the fracture
increment at time steps at each point of the front. Consequently,
many investigations have been performed tending to reveal those
general properties of the solution, which may serve for the problem
simplification. They have provided knowledge on the asymptotics of
the solution, possibility to neglect the lag between the fluid
front and the fracture contour and on the typical regimes (e.g.
\cite{Spence&Sharp}, \cite
{Desroches-et-al-1994}, \cite{GaragashDetour-2000}, \cite{AdachiDetour-2002}%
, \cite{Savitski Detour-2002}, \cite{Detournay-2004}, \cite
{BungerDetourGarag-2005}, \cite{Garagash2006}, \cite{Mitchell2007},
\cite
{KovalyshenDet2009}, \cite{Hu-Garagash-2010}, \cite{GaragDetourAdachi-2011}%
). The knowledge was incorporated in the computational codes for
practical applications (e.g. \cite{Jamamoto et al. (2004)},
\cite{Adachi-et-Al-2007}). Still, the mentioned difficulties are not
overcome, and as emphasized in the review \cite{Adachi-et-Al-2007},
there is the need "to dramatically speed up" simulators.

The recent studies tended to address this challenge
\cite{Linkov_1} -- \cite{Linkov_4}
%
have disclosed
important features of the hydraulic fracture problem, which may be
employed for enhancing the numerical simulation. They have led to
the modified formulation of the problem using:

\begin{enumerate}
\item  the particle velocity, as a variable with continuous spatial
derivative near the fluid front, instead of the pressure;

\item  the opening taken in a degree, defined by its asymptotic behavior at
the fluid front, instead of the opening itself;

\item  the speed equation (SE) at each point of the front to trace the
fracture propagation by the well-developed methods (see, e.g. \cite
{Sethian-1999}), instead of the commonly employed single equation of
the global mass balance; the speed equation also presents the basis
for proper regularization;

\item  $\varepsilon$-regularization, that is imposing the boundary condition and the
speed equation at a small distance from the front rather than on the
front itself, to exclude deterioration of the solution near the
front caused by the disclosed fact (\cite{Linkov_1}, \cite{Linkov_2})
that the boundary value problem is ill-posed when neglecting the
lag;

\item  the spatial coordinates moving with the front and evaluation of the
temporal derivative under fixed values of these coordinates;

\item  reformulation of the common system of equations and boundary
conditions in terms of the suggested variables complimented with
$\varepsilon$-regularization.
\end{enumerate}

The computational advantages of the modified formulation have been
demonstrated \cite{Linkov_1}, \cite{Linkov_2} by revisiting the classical
Nordgren \cite{Nordgren} problem. For it, common (without $\varepsilon$-regularization)\
time stepping procedures could not provide reliable third digit and they led
to strong deterioration of the solution near the fluid front for fine
meshes. In contrast, applying $\varepsilon$-regularization easily provided the solution
with relative error less than $10^{-4}$ in a time stepping procedure; the
solution was extremely stable and it never deteriorated near the front.
Using of the suggested variables made it also possible \cite{Linkov_4} to
obtain analytical solutions of the Nordgren \cite{Nordgren} and Spence and
Sharp \cite{Spence&Sharp} problems.

In this paper, we make a further step in employing the modified formulation
for (i) studying the stiffness of the system of differential equations
arising after spatial discretization, (ii) increasing the efficiency of
numerical tracing of the fracture propagation, and (iii) studying the
sensitivity. To simplify the exposition and to compare the numerical results
with benchmarks, we address the Nordgren problem.

The structure of the paper is as follows. In Section 2, we briefly
review various formulations of the problem. They include the common
formulation (Subsection 2.1), the mentioned modified formulation
(Subsection 2.2), its specification for the Nordgren problem
(Subsection 2.3) and the self-similar formulation for 1D problems.
The latter serves us to write down the benchmark analytical solution
of the Nordgren problem in the case of zero leak-off \cite{Linkov_4}
and to obtain a benchmark solution for non-zero leak-off (Subsection
2.4). Further analysis employs the \textit{modified formulation} and
the \textit{benchmark solutions}.

Section 3 presents alternative approaches to spatial discretization
of the lubrication partial differential equation (PDE) and the speed equation.
It is shown that they result in
various systems of ordinary differential equations (ODE) with quite
different stiffness. It appears that in schemes avoiding
approximation of the second spatial derivative by assuming the
particle velocity fixed at an iteration step, the stiffness, in
general, is not high being of order $O(N)$, where $N$ is the number
of nodal points. In the case of constant particle velocity, the
stiffness becomes of order $1/\varepsilon $ independently on $N$
($\varepsilon $ is the regularization parameter). This indicates
favorable features of iterative schemes with the velocity fixed at
the stage of integration in time. In contrast, in schemes employing
approximation of the second spatial derivative, the stiffness
depends on the number $N$ more strongly: it is of order $N^{2}$ for
the Nordgren problem and of order $N^{3}$ in the general case when
the net pressure is connected with the opening by exact equations of
the elasticity theory.

Section 4 presents two alternative approaches based on the
approximation of the second spatial derivative. The first of them
employs the possibility to add\ the SE to the system of ODE,
resulting from the spatial discretization
of the modified lubrication equation. 
In this way, we obtain a \emph{new well-posed} formulation for the joined system of ODE with 
\emph{initial} (Cauchy) conditions on the fluid front. The formulation opens the possibility 
to employ methods, like those of Runge-Kutta, for solving the ODE.  
It serves us to use a standard MATLAB solver in further evaluations of the accuracy and sensitivity.
The second approach employs classical
Crank-Nicolson scheme to reduce the problem to tri-diagonal
algebraic system. In this case, non-linear factors in front of the
derivatives are iterated within a time step. The SE is used at
iterations to find the new location of the fluid front. The rest of
Section 4 contains numerical results for these approaches and their
discussion. It appears that the numerical procedures resulting from
each of them are highly accurate, stable and robust. For comparison,
we also present results obtained by using the equation of the global
mass balance instead of the SE. It appears that the accuracy
becomes notably (an order, at least) less than that when using the
SE.

In Section 5 we study sensitivity of the solution to changes of influx which is one of the
major parameters of the problem. Conclusions are drawn in Section 6.

\section{Problem formulation}

\subsection{Conventional formulation}

As mentioned, a mathematical formulation of the problem includes three
groups of equations. Firstly we present them in the conventional form.

\textit{fluid equations} include the volume conservation law%
\begin{equation}
\label{e1}
\frac{\partial w}{\partial t}+\mbox{div}\,\mathbf{q}+q_{l}=0
\end{equation}%
and the relation of the Poiseuille type obtained by integration of
Navier-Stokes equations for a flow of viscous fluid in a narrow channel%
\begin{equation}
\label{e2}
\mathbf{q}=-D(w,p)\,\mbox{grad}\,p.
\end{equation}

Herein, $w(\mathbf{x},t)$ is the channel width (fracture opening), $\mathbf{q%
}(\mathbf{x},t)$ is the flux vector through the fracture height, $q_{l}(%
\mathbf{x},t)$\ is the intensity of distributed sinks or sources (below this
term will be assumed positive to account for leak-off), $p(\mathbf{x},t)$ is
the pressure, $D$ is a function or operator, such that $D(0,p)\mbox{grad}\,p=\mathbf{%
0,}$ $\mathbf{x}$ denotes the vector of the position of a point on the
surface of the flow, $t$ is the time. The flux, divergence and gradient are
defined in the tangent plane to the surface of the flow.

Substitution of (\ref{e2}) into (\ref{e1}) yields the lubrication
(Reynolds) equation

\begin{equation}
\label{e3} \frac{\partial w}{\partial
t}-\mbox{div}\,\big(D(w,p)\,\mbox{grad}\, p\big)+q_{l}=0.\quad
\end{equation}

In hydraulic fracture problems, the opening is not known in advance. Thus
its initial spatial distribution should be defined at start time $t_{0}$:

\begin{equation}
\label{e4} w(\mathbf{x},t_{0})=w_{0}(\mathbf{x}),\quad
\end{equation}%
where $w_{0}(\mathbf{x})$ is a prescribed function.

The spatial operator in (\ref{e3}) is of the second order and
elliptic. It requires only one boundary condition at the fluid
contour $L_{e}$. For instance, when neglecting the lag between the
fluid front and the fracture contour, it may be the condition of the
prescribed normal component $q_{n}$ of the flux:
\begin{equation}
\label{e5}
 q_{n}(\mathbf{x},t)=q_{0}(\mathbf{x},t),\quad \mathbf{x}\in
L_{e},
\end{equation}
 where $q_{0}(\mathbf{x},t)$ is a known function at $L_{e}$;
specifically, at the points of the fluid injection it is defined by
the injection regime; at the points of the fluid front, coinciding
with the fracture contour, we have $w=0$ and equation (\ref{e2})
implies $q_{0}(\mathbf{x},t)=0$.

In conventional formulations (e.g. \cite{Spence&Sharp}, \cite
{AdachiDetour-2002}, \cite{Savitski Detour-2002},
\cite{Detournay-2004},
\cite{Garagash2006}, \cite{Jamamoto et al. (2004)}, \cite{Adachi-et-Al-2007}%
, \cite{KovalyshenDet2009}, \cite{Hu-Garagash-2010}, \cite
{GaragDetourAdachi-2011}), to follow the fluid front propagation,
authors use the equation of the global mass balance. Being a single
equation, it do may serve for this purpose when considering 1-D
problems with one point of
the front to be traced. However in general, as emphasized in
\cite{Linkov_1} -- \cite{Linkov_4},
it is
preferable to employ the speed equation, which is formulated at each
point of the fluid front. Even for 1-D problems, it provides
advantages discussed in following sections. We shall not dwell on
this issue here as the discussion below focuses on a 1-D
problem.

\textit{Solid mechanics equations} define a dependence of the opening on the
net pressure caused by deformation of rock:%
\begin{equation}
\label{e6}
 {\cal A}w=p,\quad
\end{equation}%
with the condition of zero opening at each point $\mathbf{x}_{c}$\ of the
fracture contour:

\begin{equation}
\label{e7}
w(\mathbf{x}_{c})=0.\quad
\end{equation}

Commonly, the operator ${\cal A}$ in (\ref{e6}) is obtained by using the
theory of linear static elasticity. As mentioned, when neglecting the lag,
the condition of zero opening (\ref{e7}) replaces the condition of
zero flux on the front. Henceforth, we shall consider this case and
write $x_{c}=x_{\ast }$ with the star marking that a quantity refers
to the fluid front.

\textit{Fracture mechanics equations} define the critical state and the
perspective direction of the fracture propagation. In the commonly
considered case of the tensile mode of fracture, these are:

\begin{equation}
\label{e8}
 K_{I}(\mathbf{x}_{c})=K_{IC},\text{ \ \ \ \
}K_{II}(\mathbf{x}_{c})=0,\quad
\end{equation}%
where $K_{I}\ $is the tensile stress intensity factor (SIF), $K_{IC}$ is its
critical value,\ $K_{II}$ is the shear SIF.

The problem consists in solving the PDE (\ref{e3}) together with the
elasticity equation (\ref{e6}) under the initial condition
(\ref{e4}), boundary conditions (\ref{e5}), (\ref{e7}) and the
fracture conditions (\ref{e8}). As mentioned, the global mass
balance is usually employed instead of the speed equation to find a
current position of the front.

\subsection{Modified formulation of fluid equations and boundary conditions}

The modified formulation
\cite{Linkov_1} -- \cite{Linkov_4}
concerns mostly with the fluid
equations and corresponding boundary conditions. It employs the
primary quantity resulting from integration of the Navier-Stokes
equations when a flow occurs in a narrow channel; this is the
particle velocity. Its value $\mathbf{v}$ averaged across the
channel height defines the flux $\mathbf{q}$ entering equations
(\ref{e1}), (\ref{e2}), (\ref{e5}), because by definition

\begin{equation}
\label{e9}
 \mathbf{q}=w\mathbf{v}.
\end{equation}

Thus we may use the fluid \textit{particle velocity}%
\begin{equation}
\label{e10}
 \mathbf{v=}\frac{\mathbf{q}}{w}
\end{equation}%
instead of the flux $\mathbf{q}$. In terms of the particle velocity,
the conservation law (\ref{e1}) and the Poiseuille type equation
(\ref{e2}) become,
respectively:%
\begin{equation}
\label{e11}
 \frac{\partial w}{\partial
t}+\mbox{div}\,(w\mathbf{v)}+q_{l}=0,\quad
\end{equation}%
and%
\begin{equation}
\label{e12}
\mathbf{v}=-\frac{1}{w}D(w,p)\,\mbox{grad}\,p.
\end{equation}%
For the velocity component $v_{n}$ in a direction $\mathbf{n}$, the equation
(\ref{e12}) yields%
\begin{equation}
\label{e13}
 v_{n}=-\frac{1}{w}D(w,p)\frac{\partial p}{\partial n}.
\end{equation}

In contrast with the flux, pressure and opening, the particle velocity is a
smooth function near the fluid front. It follows from the fact that the
particle velocity $\mathbf{v(x}_{\mathbf{\ast }}\mathbf{)}$ equals to the
front speed $\mathbf{V}_{\ast }$ at each point $\mathbf{x}_{\ast }$\ of the
front. In terms of the components $v_{n\ast }$, $V_{\ast }$,\ normal to the
front, we have:%
\begin{equation}
\label{e14}
v_{n\ast }(\mathbf{x}_{\ast })=\frac{dx_{n\ast }}{dt}=V_{\ast }(\mathbf{x}%
_{\ast }),
\end{equation}%
where $x_{n\ast }$\ is the normal component of a point $\mathbf{x}_{\ast }$
on the front, $V_{\ast }=\left\vert \mathbf{V}_{\ast }\right\vert $. Hence,
the particle velocity is finite at the front in common cases of the front
propagation with a finite speed. Moreover, it is non-zero \ except for flows
with stagnation points.

The equation (\ref{e14}), where the normal component of the particle
velocity is defined by (\ref{e13}), presents the \textit{speed
equation }for the problem of
hydraulic fracture:%
\begin{equation}
\label{e15}
V_{\ast }(\mathbf{x}_{\ast })=-\frac{1}{w(\mathbf{x}%
_{\ast })}D(w,p)\frac{\partial p}{\partial n_{\ast }}.
\end{equation}%
Herein, $\mathbf{n}_{\ast }$ is the unit normal to the front in the
direction of its propagation at a point $\mathbf{x}_{\ast }$. Being the
starting concept of the theory of propagating surfaces \cite{Sethian-1999},
the speed equation is fundamental for proper tracing the hydraulic fracture
propagation.

The speed equation (\ref{e15}) yields also important implications
for numerical simulation of hydraulic fractures by finite
differences (FD). Indeed, when
at a time step we have known both $\mathbf{x}_{\ast }$ and $V_{\ast }(%
\mathbf{x}_{\ast })$, the equation (\ref{e15}) becomes a boundary
condition additional to the boundary condition (\ref{e7}) on the
front. Thus, as noted in \cite{Linkov_1}, a boundary value problem
may appear overdetermined and ill-posed in the Hadamard sense
\cite{Hadamard-1902}. To avoid difficulties, it is reasonable to use
\textit{$\varepsilon$-regularization}, suggested and successfully
used in \cite{Linkov_1}, \cite{Linkov_2}.

The $\varepsilon$-regularization is performed as follows. An exact boundary condition on
the fluid front is changed to an approximate equality at a small distance $%
r_{\varepsilon }$ behind the front. This approximate equality is
obtained by combining the boundary condition at the fluid front,
particular for a considered problem, with the speed equation, which
is quite general. In practical calculations, the distance (absolute
$r_{\varepsilon }$ or relative $\varepsilon $) is taken small enough
to use the equality sign in the derived approximate condition. This
gives us the \textit{$\varepsilon$-regularized boundary condition}
near the front. The speed equation is also assumed to be met at the
distance $r_{\varepsilon }$ with an accepted accuracy. This gives us
the \textit{$\varepsilon$-regularized speed equation. }The
$\varepsilon$-regularized boundary condition allows one to avoid the mentioned
unfavorable computational effects; the $\varepsilon$-regularized speed equation
serves to find the front propagation.

In this way, the boundary conditions (\ref{e7}) and (\ref{e1}5) are
combined to obtain the
$\varepsilon$-regularized boundary condition \cite{Linkov_2}, \cite{Linkov_3}:%
\begin{equation}
\label{e16}
 \int_{p_{0}}^{p_{\varepsilon
}}\frac{1}{w}D(w,p)dp=V_{\ast }r_{\varepsilon },
\end{equation}%
where $p_{\varepsilon }=p(r_{\varepsilon })$ is the pressure at the
distance $r_{\varepsilon }$ from the front. The
$\varepsilon$-regularized form of the speed
equation (15) is:%
\begin{equation}
\label{e17}
V_{\ast }(t)=\frac{dx_{n\ast }}{dt}=-\frac{1}{w}D(w,p)\frac{\partial p}{%
\partial n}_{r_{\varepsilon }}.
\end{equation}

The equations (\ref{e16}) and (\ref{e17}) actually employ the system
moving with the front. Thus it is reasonable to re-write the
lubrication equation (\ref{e11}) in this system. In it, the $r$-axis
is directed opposite to the front velocity, while the other axis is
tangent to the front. The connection between the temporal
derivatives evaluated under constant $x$ and $r$ is given by the
rule:

\begin{center}
$\frac{\partial }{\partial t}\big|_{\mathbf{x}=const}=\frac{\partial
}{\partial t}\big|_{r=const}+V_{\ast }\frac{\partial }{\partial r}.$
\end{center}

Then equation (\ref{e11}) reads \cite{Linkov_2}:%
\begin{equation}
\label{e18}
\frac{\partial \ln w}{\partial t}=\frac{\partial v_{n}}{\partial r}%
+(v_{n}-V_{\ast })\frac{\partial \ln w}{\partial r}-\frac{1}{w}q_{l},
\end{equation}%
where using ln$w$ serves to account for an arbitrary power asymptotic
behavior of the opening
\begin{equation}
\label{e19}
w(r,t)=C(t)r^{\alpha }+O(r^{\delta }),\quad r\to0,\quad \text{\ }\alpha \geq 0,\text{ }%
\delta >\alpha
\end{equation}%
near the front. The value of the exponent $\alpha $\ is known in a
number of important particular cases (see, e.g. \cite{Spence&Sharp}, \cite{Desroches-et-al-1994}, \cite%
{AdachiDetour-2002}, \cite{KovalyshenDet2009}), and $\delta =1+\alpha $
when the leak-off is neglected.

For the asymptotics (19), it is reasonable, in addition to the
particle
velocity, to use the variable $y(r,t)=[w(\mathbf{x}_{\ast }-r\mathbf{n}%
,t)]^{1/\alpha }$, which is linear in $r$ near the front. Finally
the lubrication equation (18) near the fluid front becomes
\begin{equation}
\label{e20}
\frac{\partial y}{\partial t}=\frac{y}{\alpha }\frac{\partial v_{n}}{%
\partial r}+(v_{n}-V_{\ast })\frac{\partial y}{\partial r}-\frac{y^{1-\alpha
}}{\alpha }\,q_{l}.
\end{equation}

In 1-D cases, the equation (\ref{e20}) is applicable to the entire
fluid. In these cases, there is the only spatial coordinate $x$ and
it is reasonable to normalize $x$ or, what is actually equivalent,
$r$ by the distance $x_{\ast }(t)$\ from the inlet to the front.
When using $\varsigma =x/x_{\ast }$, the partial derivative
evaluated under constant $r$ is expressed via that under constant
$\varsigma $\ as:
\[
\frac{\partial }{\partial t}\big|_{r=const}=\frac{\partial
}{\partial t}\big|%
_{\varsigma =const}+(1-\varsigma )\frac{V_{\ast }}{x_{\ast }}\frac{\partial }{%
\partial \varsigma }.
\]

Then in terms of $\varsigma =x/x_{\ast }=1-r/x_{\ast }$, the
lubrication equation (\ref{e20}) in 1-D cases reads:

\[
\frac{\partial \tilde y}{\partial t}=\frac{1}{x_{\ast
}}\left[ (\varsigma \tilde V_{\ast
}-\tilde v)\frac{\partial \tilde y}{\partial \varsigma }-\frac{\tilde y}{\alpha }\frac{\partial \tilde v%
}{\partial \varsigma }\right] -\frac{\tilde y^{1-\alpha }}{\alpha }\,
\tilde q_{l},
\]
where we have omitted the subscript $n$ in the notation of the
particle velocity; tilda over a symbol marks that the corresponding
function is considered to be a function of $\varsigma $:
$\tilde{y}(\varsigma ,t)=y(x_{\ast }(1-\varsigma ),t)$,
$\tilde{v}(\varsigma ,t)=v(x_{\ast }(1-\varsigma ),t)$,
$\tilde{q}_{l}(\varsigma ,t)=q_{l}(x_{\ast }(1-\varsigma ),t)$. From
now on, to simplify notation, we shall omit the tilda over functions
depending on $\varsigma $. Thus, the previous equation is written
as:

\begin{equation}
\label{e21}
 \frac{\partial y}{\partial t}=\frac{1}{x_{\ast
}}\left[ (\varsigma  V_{\ast
}- v)\frac{\partial  y}{\partial \varsigma }-\frac{ y}{\alpha }\frac{\partial  v%
}{\partial \varsigma }\right] -\frac{ y^{1-\alpha }}{\alpha }\, q_{l},
\end{equation}

Note that when $q_{l}$ near the front\ decreases faster than $w=y^{\alpha }$%
, we may divide (21) by $y$, obtaining the equation

\begin{equation}
\label{e22}
\frac{1}{y}\frac{\partial  y}{\partial t}=\frac{\varsigma  V_{\ast }- v}{%
x_{\ast } y}\frac{\partial  y}{\partial \varsigma }-\frac{1}{\alpha x_{\ast }}%
\frac{\partial v}{\partial \varsigma }-\frac{1}{\alpha y^{\alpha }}\,
q_{l}.
\end{equation}%
In (\ref{e22}), under the assumed asymptotics of $ q_{l}$, the term
$(\partial  y/\partial t)/ y$, the factor ($\varsigma  V_{\ast }-
v)/(x_{\ast } y)$ and the derivative $\partial  v/\partial \varsigma
$ are finite at the fluid front.

\subsection{Nordgren problem in modified formulation}

The 1-D problem (Fig.~1) studied by Nordgren \cite{Nordgren} is
similar to that considered by Perkins and Kern \cite{Perkins},
improving the model of these authors by rigorous mathematical
formulation, which includes finding the fracture length $x_{\ast }$
as a part of the solution. Below we use the rigorous formulation by
Nordgren and attribute it to this author.

The fluid equations of the problem are (\ref{e1}) - (\ref{e3}) with
the operator $D$ being
the multiplier%
\begin{equation}
\label{e23}
 D( w,p)=k_{l} w^{3},
\end{equation}%
corresponding to the flow of Newtonian fluid in a narrow channel with an
elliptic cross-section; for it $k_{l}=1/(\pi ^{2}\mu )$, where $\mu $\ is
the dynamic viscosity.

The operator ${\cal A}$ in the solid mechanics equation (\ref{e6}) is also
taken in the simplest form as the multiplier $k_{e}=(2/\pi h)E/(1-\nu ^{2})$ in the
linear dependence between the pressure and opening%
\begin{equation}
\label{e24}
  p=k_{e} w.
\end{equation}

\vspace*{-5mm}

\begin{figure}[h!]
\centering
    \includegraphics [scale=0.41]{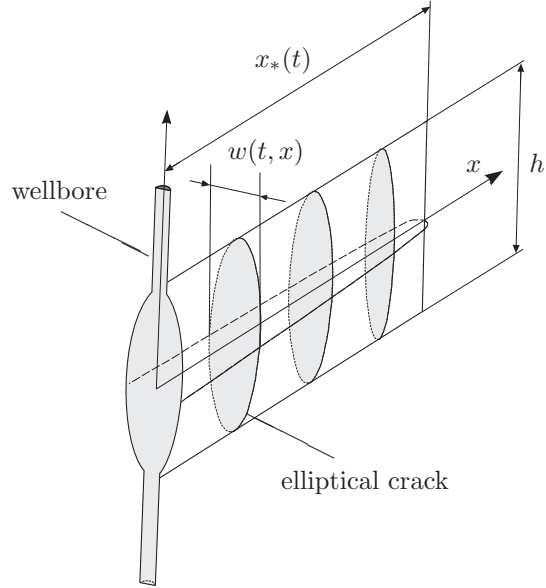}

\begin{picture}(0,0)(70,100)

\put(22,295){$w(t,x)$}

\put(112,290){$x$}

\put(136,290){$h$}

\put(-41,275){\line(2,-1){33}}
\put(-60,280){wellbore}

\put(62,182){\line(-2,1){33}}
\put(42,170){elliptical crack}

\put(32,330){$x_*(t)$}
\end{picture}

\vspace{-10mm}

\caption{Scheme of the Nordgren problem.}
\label{Fig.1}

\end{figure}

Herein, $h$ is the vertical length of a narrow elliptic channel
(Fig.~1), $E$ is the Young modulus, $\nu $ is the Poisson's ratio of
rock. For the dependence (\ref{e24}), there is no need in a fracture
criterion because it does not involve the fracture front.

The initial condition (\ref{e4}) in the 1-D case reads
\begin{equation}
\label{e25} w(x,t_{0})= w_{0}(x),\quad
\end{equation}
with $ w_{0}(x)=0$ ahead of the fluid front $x_{\ast}(t_0)$.

In view of (\ref{e2}), (\ref{e23}) and (\ref{e24}), the boundary
condition (\ref{e5}) of the prescribed influx $ q_{0}$\ at the inlet
$x=0$ becomes:
\begin{equation}
\label{e26} -\frac{k_{l}k_{e}}{3} w(0,t)\frac{\partial
w^{3}}{\partial x}\big|_{x=0}= q_{0}(t).
\end{equation}
The boundary condition of zero opening involves the only point
$x_{\ast}=x_*(t)$ of the fluid front:
\begin{equation} \label{e27}
w(x_*)=0.\quad
\end{equation}
The equation for the particle velocity (\ref{e12}) and the speed
equation (\ref{e15}) become, respectively,
\begin{equation}
\label{e28}
  v=-\frac{k_{l}k_{e}}{3}\frac{\partial w^{3}}{\partial x}
\end{equation}
and
\begin{equation}
\label{e29}
 V_{\ast }=\frac{dx_{\ast
}}{dt}=-\frac{k_{l}k_{e}}{3}\frac{\partial  w^{3}}{
\partial x}\big|_{x=x_*(t)}.
\end{equation}

The exponent
$\alpha $ in (19) equals 1/3 (see, e.g. \cite{KovalyshenDet2009})
when the leak-off is neglected or even when it is described by the
Carter's dependence and, consequently, singular at the crack tip.
Then $y=w^{3}$, and in terms of the normalized independent variable
$\varsigma$, the modified equation (21) for the 1-D problem (21)
reads:
\begin{equation}
\label{e30}
 \frac{\partial  y}{\partial t}=\frac{1}{x_{\ast
}} \left[ (\varsigma V_{\ast }- v)\frac{\partial  y}{\partial \varsigma }-3 y\frac{\partial  v}{%
\partial \varsigma }\right] -3 y^{2/3}  q_{l},
\end{equation}
where the particle velocity $v$ is connected with $y$
by equation following from (\ref{e28}):
\begin{equation}
\label{e31}
 v=-\frac{k_{l}k_{e}}{3x_{\ast }}\frac{\partial
 y}{\partial \varsigma }.
\end{equation}
 In the new variables, the asymptotic equation (19) with $\alpha =1/3$ is
written as $y=C^{3}(t)x_{\ast }(1-\varsigma )+O((1-\varsigma
)^{1+\kappa })$ for $\varsigma \rightarrow 1$, where $\kappa >0$ for
any leak-off tending to
zero at the crack tip. Then (29) and (31) imply that near the front $%
v=V_{\ast }=\frac{k_{l}k_{e}}{3}C^{3}(t).$ Hence, the coefficient $C^{3}(t)$%
\ is a multiple of the front speed:
$C^{3}(t)=\frac{3}{k_{l}k_{e}}V_{\ast }(t)$, and the asymptotics of
$y$ near the front is
\begin{equation}
\label{e32}
 y(\varsigma)=  \frac{3}{k_{l}k_{e}} V_{\ast
}(t)x_{\ast }(t)(1-\varsigma )+O\big((1-\varsigma)^{1+\kappa}\big),
\quad \varsigma\to1.
\end{equation}

In terms of the variables $ v$, $ y$ and $\varsigma $, the
conditions (\ref{e25}) and (\ref{e26}) read, respectively:
\begin{equation}
\label{e33}
  y(\varsigma ,t_{0})= y_{0}(\varsigma ),\quad
\end{equation}%
\begin{equation}
\label{e34}
 \sqrt[3]{y(0,t)}v(0,t)= q_{0}(t),
\end{equation}
where $ y_{0}(\varsigma )=w_{0}^{3}(\varsigma x_{\ast })$ for $0\leq
\varsigma \leq 1$ and $ y_{0}(\varsigma )=0$ ahead of the fluid front ($%
\varsigma >1$).

From (\ref{e32}), (\ref{e29}) it follows that the
$\varepsilon$-regularized boundary condition (\ref{e16}) and the
$\varepsilon$-regularized speed equation (\ref{e17}) are,
respectively:
\begin{equation}
\label{e35}
 y(t,1-\varepsilon )=\frac{3}{k_{l}k_{e}}
V_{\ast }x_{\ast }\varepsilon ,
\end{equation}%
\begin{equation}
\label{e36}
  V_{\ast }x_{\ast
}=-\frac{k_{l}k_{e}}{3}\frac{\partial  y}{\partial \varsigma
}\Big|_{\,\varsigma =1-\varepsilon},
\end{equation}
where $\varepsilon $ is a small relative distance from the front ($%
\varepsilon =r_{\varepsilon }/x_{\ast }$) and $ V_{\ast }=dx_{\ast
}/dt$.

We need to solve the PDE (\ref{e30}), where $ v$ is connected with $
y$ by equation (\ref{e31}), under the initial condition (\ref{e33}),
the boundary condition (\ref{e34}) at the inlet and the
$\varepsilon$-regularized boundary condition (\ref{e35}) imposed at
a small relative distance $\varepsilon $ from the fluid front. The
regularized speed equation (\ref{e36}) serves to find the position
of the fluid front.

Emphasize that, considering BVP, we do not use the conditions (\ref{e27}) and
(\ref{e29}), not involving regularization, to avoid deterioration of
the solution near the front.

\subsection{Self-similar formulation. Benchmark solutions}

As shown by Spence and Sharp \cite{Spence&Sharp}, 1-D plane and axisymmetric
problems may be reduced to a self-similar formulation in the case when there
is no leak-off and the flux $q_{0}$\ at the inlet is proportional to a power
$\varphi (\beta ,t)=t^{\beta }$\ or exponential $\varphi (\beta ,t)=e^{\beta
t}$ function of time with constant $\beta $. In the particular case of
constant influx, $\beta =0$. Representing a solution in the form $%
 w(t,\varsigma )=\varphi (\beta _{w},t)W(\varsigma )$, $ p(t,\varsigma
)=\varphi (\beta _{p},t)P(\varsigma )$ with separated temporal
$\varphi (\gamma ,t)$ and spatial $\varsigma =x/x_{\ast }$ variables
leads to equations with the only independent variable $\varsigma $
and with $x_{\ast }=B\varphi (\beta _{\ast },t)$. For an
axisymmetric problem, $x_{\ast }$ is a current radius of the
fracture. The constants $\beta _{w}$, $\beta _{p}$ and $\beta _{\ast
}$ depend on a particular 1D problem and $\beta $.

Actually, Nordgren employed this option for the case of constant influx ($%
\beta =0$) (\cite{Nordgren}, Appendix C). We shall use the separation of
variables with $\varphi (\beta ,t)=t^{\beta }$ to find benchmark solutions
needed for further discussion. We include the case of non-zero leak-off by
representing the leak-off term in separated variables, as well.

In this way, for the
considered problem, the speed equation yields $\beta _{\ast
}=(3\beta _{w}+1)/2$, $B=\sqrt{V(1)/\beta _{\ast }}$, while the PDE
(\ref{e30}) becomes the ordinary differential equation (ODE):
\begin{equation}
\label{e37}
\frac{d V}{d\varsigma }+\frac{ V(\varsigma )-\varsigma   V(1)}{3Y}\frac{dY}{%
d\varsigma }+\beta _{w}+Y^{-1/3}Q_{l}(\varsigma )=0,
\end{equation}%
where $  V(\varsigma )$ is the self-similar particle velocity defined as $%
  V(\varsigma )=-\frac{k_{l}k_{e}}{3}\frac{dY}{d\varsigma }$
with $Y(\varsigma )=W^{3}(\varsigma )$. For consistency, the leak-off term $ q_{l}$ entering
(\ref{e30}) is also taken as a function in separated variables $ q_{l}=Q_{l}(%
\varsigma )t^{\beta-1}$.

The condition (\ref{e34}) at the inlet defines the dependence of
$\beta _{w}$ on the exponent $\beta $\ in the influx prescribed by
$ q_{0}=At^{\beta }$, where $A$ is a constant: $\beta
_{w}=(1+2\beta )/5$. Then $\beta_{\ast }=(3\beta +4)/5$. Noting that
the front propagates with the speed $ V_{\ast }(t)=B\beta _{\ast }t^{\beta _{\ast }-1}$, this implies that the
propagation speed is constant when $\beta _{\ast }=1$, what
corresponds to $\beta =1/3$. For a constant influx ($\beta =0)$, we
have the Nordgren's results: $\beta _{w}=1/5$, $\beta _{\ast }=4/5$
and the propagation speed changes proportionally to $t^{-1/5}$.

\subsubsection{Benchmark solution for zero leak-off}

For zero leak-off $q_{l}=0$, (or $Q_{l}(\varsigma )=0$) and constant influx ($%
\beta =0$, $\beta _{w}=1/5$, $\beta _{\ast }=4/5$) the self-similar
formulation serves to obtain the analytical solution \cite{Linkov_4}. In
terms of physical quantities it is:

\begin{equation}
\label{self-sim}
\begin{array}{c}
 w(\varsigma,t)=w_{n}\sqrt[3]{Y(x/x_{\ast })}t^{1/5}, \quad  y=y_{n}Y(x/x_{\ast })t^{3/5}, \quad p=k_{e}w,\\[2mm]
 v=v_{n}v_{\psi }(x/x_{\ast })t^{-1/5}, \quad  V_{\ast }(t)=0.8\xi _{\ast
}v_{n}t^{-1/5}, \quad x_{\ast }(t)=\xi _{\ast }x_{n}t^{4/5}.
\end{array}
\end{equation}

Herein, $x_{n}=(k_{l}k_{e}/4)^{1/5}q_{n}^{3/5}t_{n}^{4/5}$, $%
w_{n}=q_{n}t_{n}/x_{n}$, $y_{n}=w_{n}^{3}$, $v_{n}=x_{n}/t_{n}$, $q_{n}$,
and $t_{n}$ are \textit{normalizing} length, opening, cubed opening,
particle velocity, flux and time, respectively. The normalizing quantities $%
q_{n}$, $t_{n}$ may be chosen as convenient. The dimensionless parameter $%
\xi _{\ast }$\ is defined by the prescribed influx $q_{0}$ at the inlet \cite%
{Linkov_2}:

\begin{center}
$\xi _{\ast }=1.3208446(q_{0}/q_{n})^{0.6}$.
\end{center}

Thus, when taking the influx $ q_{0}$ as the normalizing flux $q_{n}=q_{0}$, one has $\xi _{\ast }=1.3208446$. As above, $\varsigma =x/x_{\ast }$ is
the relative distance from the inlet. The functions $v_{\psi }(\varsigma )$
and $Y(\varsigma )$ are given by the series:%
\begin{equation}
\label{e38}
v_{\psi }(\varsigma )=0.8\xi _{\ast
}\sum_{j=0}^{j=\infty }b_{j}(1-\varsigma
)^{j},\text{ \ }Y(\varsigma )=0.6\xi _{\ast }^{2}\sum_{j=1}^{j=\infty }\frac{%
b_{j-1}}{j}(1-\varsigma )^{j},
\end{equation}%
where $b_{0}=1$, $b_{1}=-1/16$, and the next coefficients are found
recurrently as
\[
b_{j+1}=-\frac{1}{3j+4}\left[ \frac{4j+1}{4(j+1)}\,b_{j}+\sum_{k=2}^{j+1}\frac{%
3j-2k+6}{k}b_{k-1}b_{j-k+2}\right],\quad j=1,2,... .
\]

The series (39) rapidly converge. Five first terms with coefficients $%
b_{0}=1 $, $b_{1}=-1/16$, $b_{2}=-(15/224)b_{1}$, $b_{3}=-(3/80)b_{2}$, $%
b_{4}=-(11/5824)b_{3}$ provide the accuracy of seven significant
digits in
the entire interval of flow. The corresponding relative error is of order $%
10^{-5}$ even near the fluid front where $Y(\varsigma )\rightarrow
0$. In further calculations, we use seven terms what guaranties that
the relative error is of order $10^{-7}$. For $\xi _{\ast }=1$, the
normalized self-similar particle velocity $v_{\psi }(\varsigma
)=V(\varsigma )$\ and the cubed normalized opening $Y(\varsigma )$
present the solution of the equation (37), when the variables in
the latter are normalized in accordance
with (38). Then $Y(1)=0$, $V(1)=0.8$, $V(\varsigma )=-\frac{4}{3}\frac{dY}{%
d\varsigma }$ and $Q_{l}(\varsigma )=0$. Below we shall use the
solution
(38), (39) with $q_{n}=q_{0}=1$, $t_{n}=1$, $k_{l}=1$, $k_{e}=1$; then $%
x_{n}=4^{-1/5}$, $w_{n}=4^{1/5}$, $\xi _{\ast }=1.3208446$. We shall
call it the \textit{benchmark solution I}.

\subsubsection{Benchmark solution for non-zero leak-off}

In this case, we prescribe the function $W(\varsigma )$ and define
the corresponding leak-off term $Q_{l}(\varsigma )$ entering
(\ref{e37}) as such, for which the lubrication equation is satisfied
by $W(\varsigma )$. Specifically, for a prescribed function
$W(\varsigma )$, the latter
satisfies (\ref{e37}), when assuming $Q_{l}(\varsigma )=$ $-Y^{1/3}\left[ \frac{d  V%
}{d\varsigma }+\frac{ V(\varsigma )-\varsigma
 V(1)}{3Y} \frac {dY}{d \varsigma}+\beta _{w}\right] $. The initial
condition (\ref{e33}) becomes $ y(\varsigma
,t_0)=t_{0}^{3\beta_w }W^3(\varsigma)$. We specify the function
$W(\varsigma )$ by the expression:
\begin{equation}
\label{e39}
W(\varsigma )=\omega (1-\varsigma
)^{1/3}[1+s(\varsigma)],
\end{equation}%
where $\omega $ is a constant, $s(\varsigma )=C_{W}(1-\varsigma
)+O((1-\varsigma )^{2})$ as $\varsigma \rightarrow 1$, and $C_{W}$
is a constant chosen in such a way that the particle velocity and
leak-off term are positive in the entire flow region. The choice of
$W(\varsigma )$ in the form (40) guaranties the asymptotic behavior
(19) of the crack opening.

The benchmark solution, corresponding to the choice (40), is:

\[
 w(\varsigma,t)=W(\varsigma)t^{\beta_w}, \quad  y= w^{3},
\quad p=k_{e}w,\quad  \beta _{w}=(1+2\beta )/5, \quad \beta_{\ast
}=(3\beta +4)/5,
\]
\[
 v(\varsigma,t)=\sqrt{\frac{3k_e
k_l \omega^3(3\beta+4)}{5}}\,t^{\frac{3\beta-1}{5}}\Big(\frac{1}{3}
[1+s(\varsigma)]-(1-\varsigma)[1+s(\varsigma)]s'(\varsigma)\Big),
\]
\[
q_0(t)=-\sqrt{\frac{3k_ek_l\omega (3\beta+4)}{5}}\omega^2
t^\beta\,[1+s(0)]^3\{s'(0)-[1+s(0)]/3\},
\]
\[
 V_{\ast }(t)=\sqrt{\frac{k_e k_l
\omega^3(3\beta+4)}{15}}\,t^{\frac{3\beta-1}{5}}, \quad x_{\ast
}(t)=\sqrt{\frac{5k_ek_l\omega^3}{9\beta+12}}\,t^{\beta_*},\]
\[
  q_l=\frac{t^\frac{2\beta-4}{5}}{3W^2(\varsigma)}\left(\frac{3\beta+4}{15\omega^3}\left[\left(\frac{\partial W^3}{\partial \varsigma}-\varsigma\frac{\partial W^3}
 {\partial \varsigma}\big|_{\varsigma=1}\right) \frac{\partial W^3}{\partial \varsigma} +3\frac{\partial^2W^3}{\partial \varsigma^2} \right]
 -\frac{6\beta +3}{5}W^3\right).
\]

The initial condition (25) reads $w(x,t_{0})=W(x/x_{\ast
})t_{0}^{\beta _{w}} $. For certainty, in further calculations we
set $t_{0}=1$, $k_{e}=1$, $k_{l}=1$, $\omega =1$ and consider the
case of constant influx: $\beta =0$ ($\beta _{w}=1/5$, $\beta _{\ast
}=4/5$). Below we shall use two choices
of the function $s(\varsigma )$ with the same $C_{W}=-1/(96e)$. One of them $%
s(\varsigma )=-(96e)^{-1}(1-\varsigma )$ corresponds to small leak-off; it
will be referred as the \textit{benchmark solution II}. The other $%
s(\varsigma )=-(96e)^{-1}(1-\varsigma )$ $+0.05(1-\varsigma )^{2}$
describes
notable leak-off; it will be referred as the \textit{benchmark solution III}.

In order to compare the benchmark solutions I, II and III, we
evaluate the
total fluid loss per unit time at the initial moment $t_{0}=1$: $%
Q_{l}=\int_{0}^{1}q_{l}(\varsigma ,1)d\varsigma $. For the benchmark
solution I, we have $Q_{l}=0$; for the benchmark solution II, the
total loss is $Q_{l}=3.3\cdot 10^{-3}$, $Q_l/q_0=4\cdot 10^{-3}$;
for the benchmark solution III, the total loss becomes an order greater being $Q_{l}=4.8\cdot
10^{-2}$, $Q_{l}/q_0=4\cdot 10^{-2}$. They also correspond to different distributions of the
particle velocity in the flow region what has strong impact on the
accuracy of calculations. We may characterize the variation of the
velocity by the parameter

\begin{center}
$\gamma _{v}=[\max (v(\varsigma ,t))-\min (v(\varsigma ,t)]\left[
\int_{0}^{1}v_{l}(\varsigma ,t)d\varsigma \right] ^{-1}$.
\end{center}

Its values are: $\gamma _{v}=0.06$ for the benchmark solution I,
$\gamma _{v}=0.02$ for the benchmark solution II, and $\gamma
_{v}=0.4$ for the benchmark solution III. Thus, the velocity
distribution is almost uniform for the benchmark solution II, and it
is strongly non-uniform for the benchmark solution III; the
benchmark solution I with zero leak-off presents an intermediate
case.

\section{Spatial discretization. Stiffness analysis}

Henceforth, in accordance with $\varepsilon $-regularization, we
consider the spatial interval $[0,1-\varepsilon ]$ instead of
$[0,1]$ to avoid computational difficulties disclosed in
\cite{Linkov_1}, \cite{Linkov_2}. Normally we shall set $\varepsilon
=10^{-4}$ what guaranties that the \textit{relative}
error of the cubed opening is of order $10^{-4}$ even near the fluid
front. The spatial discretization of the problem employs
representation of the interval $[0,1-\varepsilon ]$\ with $N-1$
segments of the equal length $h_\varepsilon=(1-\varepsilon )/(N-1)$. The nodal
points are $\varsigma _{j}=j(1-\varepsilon )/(N-1)$ ($j=1,2,...,N$).
Then applying a finite difference approximation (say, the left-hand side
approximation) to the spatial derivative(s) of the
unknown function $y$, the PDE (21) yields a system of ODE in the vector $%
\mathbf{Y}$ of nodal values $y_{j}$.

\subsection{Iterations in particle velocity}

Let us employ the fact that the particle velocity $v$, being the
ratio of the flux and opening, which both decrease when approaching
the fluid front, changes significantly less than these quantities.
(As established in \cite {Linkov_4}, the particle velocity is almost
constant in the entire flow region in problems of Nordgren and
Spence \& Sharp when neglecting the lag). Let us assume as a rough
estimate $v=V_{\ast }$. Then the ODE has the linear form:
\begin{equation}
\ \frac{d\mathbf{Y}}{dt}\mathbf{=A(}t\mathbf{)Y+B(}t\mathbf{)},\quad
t>t_{0},
\end{equation}
where

\begin{equation}
\ \mathbf{A(}t\mathbf{)=}\frac{V_{\ast }}{x_{\ast }}\mathbf{DE},
\end{equation}
$\mathbf{D}$ and $\mathbf{E}$ are diagonal and two-diagonal
matrices, whose non-zero entries are defined, respectively, as:
$d_{ii}=1-\varepsilon _{i}$,
($i=1,...,N$) and $e_{ij}=1$ for $i=j$, $e_{ij}=-1$ for $i=j+1$, ($%
i,j=1,...,N$). The eigenvalues $\lambda _{j}$ of the matrix $\mathbf{A(}t%
\mathbf{)}$, defined by (42), are $\lambda _{j}=-d_{jj}\frac{V_{\ast }(t)}{%
x_{\ast }(t)}$, ($j=1,2\ldots ,N$). All of them are negative, and
the stiffness (condition) ratio $k_{A}=\frac{\max (-\lambda _{j})}{\min
(-\lambda _{j})}$, characterizing the stiffness of the system of ODE
(41), is
\begin{equation}
\kappa _{A}=\frac{d_{11}(\varepsilon )}{d_{NN}(\varepsilon )}=\frac{1}{%
\varepsilon }.
\end{equation}
Actually, it is possible to set $\varepsilon =0$ when obtaining the
system of ODE\ (41), because the initial (Cauchy) problem of solving
it under the condition of zero opening $Y(1)=0$ is well posed.
Still, effectively, when
employing the condition $Y(1)=0$, we arrive at the same system (41) with $%
N-1 $ unknowns and $\varepsilon =1/(N-1)$. Then (43) becomes
\[
\kappa _{A}=N-1.
\]

In a general case, the particle velocity is not constant along the
fracture. Nevertheless, the stiffness ratio for large $N$ can be estimated as
\begin{equation}
\label{estim}
\kappa _{A}\simeq\frac{1}{\varepsilon}\,c_{A},
\end{equation}
where $c_{A}=c_A(t)$ depends on the distribution of the
velocity along the fracture and can be computed as
\[
c_A=v(0,t)\left(V_*(t)+\frac{\partial v}{\partial\varsigma}(1,t)\right)^{-1}.
\]
For the benchmark solutions under consideration, we have
the stiffness ratio (\ref{estim}) independent of time $t$: $c_A^{(I)}=0.52$, $c_A^{(II)}=0.96$, $c_A^{(III)}=1.43$.
Finally, employing the condition $Y(1)=0$, the stiffness ratio is linear in $N$.

Linear dependence on $N$ is more favorable for solving ODE, than
quadratic or cubic dependencies, which appear in other
approaches. Even for $N\sim 10^{5}$, the stiffness is of order
$10^{5}$,\ what is quite acceptable in practical calculations. This
indicates that solving the problem by iterations with the particle
velocity, found at the stage of temporal integration, may be
reasonable. However, to the moment, it is unclear how to properly organise the iterations to meet the BC of prescribed influx at the inlet.

\subsection{Spatial discretization with reduction to dynamic system of ODE}

Another way of solving the problem may consist in substitution of
the velocity, defined by (31), into (30) and considering the PDF
with the second partial derivative. The substitution may be employed
either for all terms including the particle velocity, or only for
the term $\partial v/\partial \varsigma $. Herein we employ the
first option and obtain:
\begin{equation}
\frac{\partial y}{\partial t}=\frac{k_{l}k_{e}}{x_{\ast }^{2}}\left\{ y\frac{%
\partial ^{2}y}{\partial \varsigma ^{2}}+\frac{1}{3}\left[ \frac{\partial y}{%
\partial \varsigma }-\varsigma \frac{\partial y}{\partial \varsigma }(1,t)%
\right] \frac{\partial y}{\partial \varsigma }\right\}
-3y^{2/3}q_{l}.
\end{equation}

We compliment (45) with the regularized SE (36), which, as
mentioned, serves to find the fracture length $x_{\ast }$. When
employing (45), it is extremely beneficial to re-write the SE (36)
in the form, which similarly to (45) contains $x_{\ast }^{2}$:
\begin{equation}
\frac{dx_{\ast }^{2}}{dt}=-2\frac{k_{l}k_{e}}{3}\frac{\partial
y}{\partial \varsigma }\big |_{\varsigma =1-\varepsilon }.
\end{equation}

Then after a spatial discretization and using the boundary
conditions at the inlet (34) and on the front (35), we arrive at a
system of ODE in unknown values at nodes inside the flow region and
the additional unknown $x_{\ast }^{2}$. The initial conditions for
the system are given by (33) and $x_{\ast }^{2}(t_{0})=x_{\ast
0}^{2}$. This opens the possibility to utilize well-developed
methods for solving systems of ODE. Specifically, the Runge-Kutta
method become available. Below we extensively use the new option for
studying the accuracy and sensitivity. Emphasize that it has
appeared only due to employing the (local) \textit{speed equation}
as the basis for tracing the front propagation. There is no such an
opportunity when tracing the front in conventional ways (e.g.
\cite{Jamamoto et al. (2004)}, \cite {Adachi-et-Al-2007}) by using
the \textit{global mass balance}. Below we shall also see that the
accuracy of calculations, based on the SE in the form (46), is
significantly (more than an order) better than that obtained with
the global mass balance.

We may estimate the stiffness of the dynamic system (41),
corresponding to (45) after spatial discretization. The system is
non-linear but taking into account the asymptotic behavior (32) and
the leading term of the equation (45), we obtain the following
expression for the matrix $\mathbf{A}(t)$:
\[
\mathbf{A}(t)=\frac{V_{\ast }(t)}{x_{\ast }(t)}\mathbf{EDE}^{T},
\]
where the matrices $\mathbf{D}$\ and $\mathbf{E}$ are defined above.
In this case, it is possible to evaluate the condition number
$k_{\infty }=\|{\bf A}\| _{\infty }\cdot \|
{\bf A}^{-1}\| _{\infty }$ in the space $l_{\infty }$
by using results from \cite{Higham-1986}.
When taking $\varepsilon =1/(N-1)$, the
estimation for large $N$ is:
\[
k_{\infty }=2N^{2}(\ln N+\gamma -1)+O(N),\quad
N\rightarrow \infty,
\]
where $\gamma $ is the Euler-Mascheroni constant.

\begin{figure}[h!]

\centering
\includegraphics [scale=0.40]{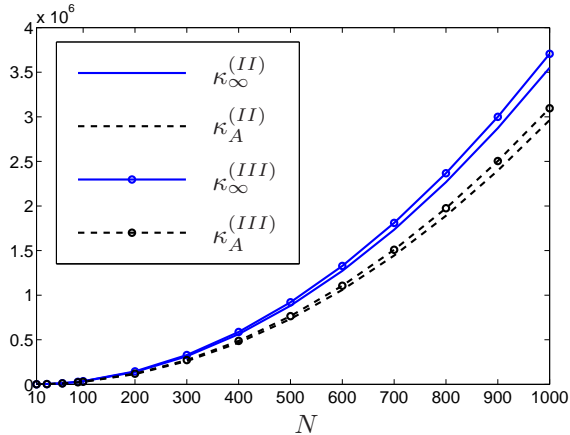}
       \put(-120,0){$N$}
    \put(-151,132){$\kappa^{(II)}_\infty$}
    \put(-151,113){$\kappa^{(II)}_A$}
    \put(-151,92){$\kappa^{(III)}_\infty$}
    \put(-151,72){$\kappa^{(III)}_A$}

    \caption{Condition number $\kappa_\infty(N)$ and stiffness ratio $\kappa_A(N)$ for the benchmark solution II (lines without markers) 
    and III (lines with markers).}

\label{cond}
\end{figure}

The dependence is close to quadratic in $N$ what means much stronger
growth of stiffness in the considered approach than in the previous
one. Again, in a general case, there is no analytical formula for the
stiffness ratio. It may be estimated as
\begin{equation}
\kappa_A=c_{A}N^{2}+O(N),\text{ \ \ }N\rightarrow \infty ,
\end{equation}
where $c_{A}$ is a constant depending on a particular problem.  We
calculated the condition number $\kappa_\infty$ and the stiffness ratio $\kappa_A$ 
numerically at $t_{0}=1$ for the
benchmark solutions I, II and III with $\varepsilon =1/(N-1)$.
The dependencies of $k_{\infty }$
and $k_{A}$ on $N$ for the benchmark solutions II and III are presented in
Fig. \ref{cond}. The graphs for the benchmark I are
indistinguishable from those for the benchmark II. It can
be seen that $k_{\infty } $\ and $k_{A}$ comply with the asymptotic estimation (47).

Note that the quadratic dependence (47) on the number of nodes $N$\
follows from the proportionality of the pressure to the opening
accepted in the Nordgren model. In the case when the pressure and
the opening are connected by the exact equations of the elasticity
theory, the dependence becomes cubic (see, e.g.
\cite{Adachi-et-Al-2007}). For $N=10^{5}$, the stiffness ratio
becomes of order $10^{15}$ what is critical for most of the solvers.
In this respect, employing iterations in the fluid velocity may be a
reasonable strategy despite it requires repeated solving of ODE and
accurate evaluation of the velocity after an iteration. Still,
\textit{using the system (45), (46) looks beneficial in a vicinity
of the fluid front} in the general case of a 2D fracture. Then the
number of nodal points may be taken small enough to avoid too stiff
system.

\section{Accuracy of computations}

We shall use two mentioned approaches for approximation of the
spatial derivatives. 

The first one, leading to the dynamic system (45), (46), has been suggested and explained when analyzing stiffness. 
It results in a \emph{well-posed initial} (Cauchy) problem for the system of ODE what opens the possibility to solve 
the problem by well-established methods, like those of Runge-Kutta. 
Making use of this opportunity, we applied the standard MATLAB solver \textit{ode15s}.
It is based on the Runge-Kutta method and employs
automatic choice of the time step.

The other option consists in substitution of the velocity, defined
by (31), into (30) only for the term $\partial v/\partial \varsigma
$. Then the equation contains the propagation speed $V_{\ast }$ and
the\ particle velocity $v$ in the multiplier by the first spatial
derivative:
\begin{equation}
\frac{\partial y}{\partial t}=\frac{k_{l}k_{e}}{x_{\ast }^{2}}y\frac{%
\partial ^{2}y}{\partial \varsigma ^{2}}+\frac{1}{x_{\ast }}(\varsigma
V_{\ast }-v)\frac{\partial y}{\partial \varsigma }-3y^{2/3}q_{l}.
\end{equation}

We compliment (48) with the regularized speed equation (36), which
after integration in time and accounting for the asymptotics (32)
may be written as
\begin{equation}
x_{\ast }^{2}=x_{\ast 0}^{2}+2\frac{k_{l}k_{e}}{3}\int_{t_{0}}^{t}
\lim_{\varsigma \rightarrow 1}\frac{y(\varsigma ,t)}{1-\varsigma
}dt.
\end{equation}

This form of equations is convenient for employing an implicit time
stepping method, for example, the classical Crank-Nicolson scheme.
The boundary conditions on a time step are obtained by using the
discretized equations (34) (at the inlet) and (35) (at the
$\varepsilon $-regularized opening condition). The resulting
non-linear algebraic system is solved by successive iterations. At
each of the iterations, we consider a linear system by taking fixed
values of $y$, $v$, $V_{\ast }$ and $x_{\ast }$ in the non-linear
terms. At an iteration, for fixed coefficients in front of the
spatial derivatives and leak-off term and for fixed coefficients in
non-linear boundary conditions, the algebraic system is
tri-diagonal. It is efficiently solved by the sweep-method. At the
end of an iteration, we
obtain new nodal values of $y$, which serve to evaluate new values of $%
V_{\ast }$ and $v$ by using (31), and new value of $x_{\ast }$ by
using (49). The non-linear terms are iterated within a time step
until the difference of values obtained on successive iterations
becomes less than a prescribed tolerance. For a sufficiently small
time step, the values obtained on the previous time step may serve
as acceptable approximations, then one iteration is usually sufficient to
meet the tolerance.

\begin{figure}[h!]

    \hspace{-2mm}\includegraphics [scale=0.40]{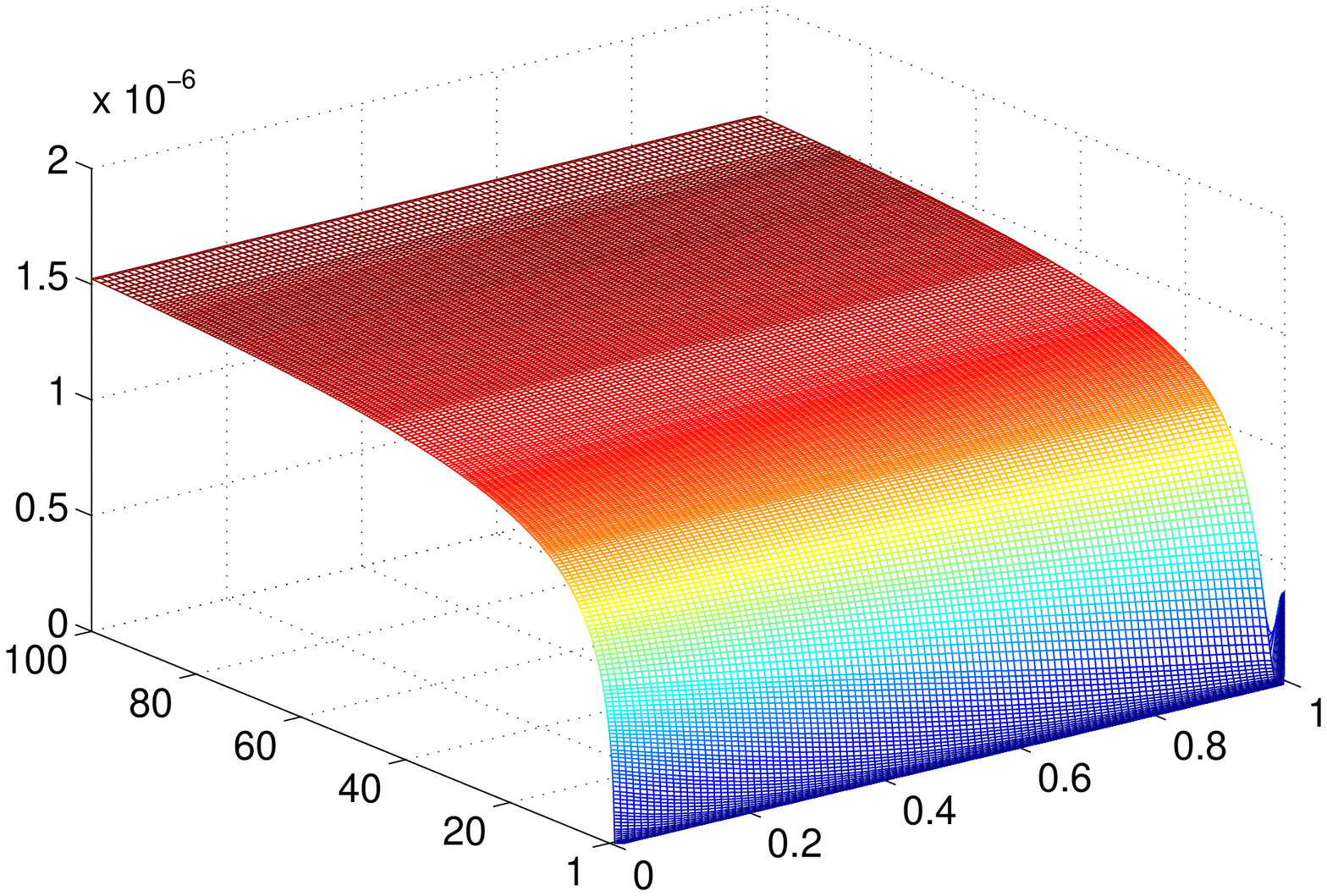}
    \put(-195,17){$t$}
    \put(-67,7){$\varsigma$}
    \put(-245,90){$\delta  w$}
    \hspace{-4mm}\includegraphics [scale=0.40]{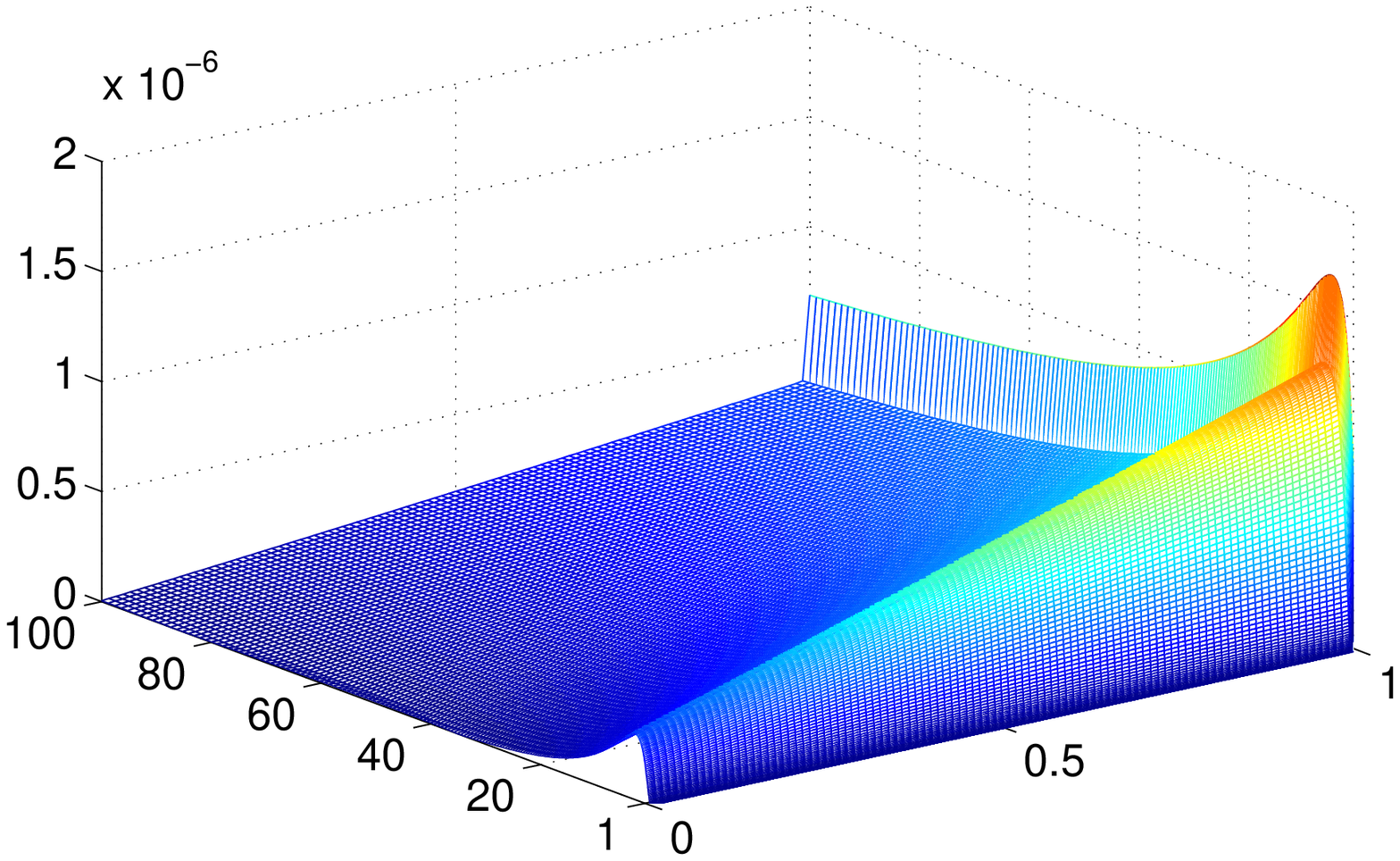}
    \put(-180,12){$t$}
    \put(-67,7){$\varsigma$}
    \put(-230,75){$\delta  w$}

    \caption{Relative error of the crack opening for the benchmark solution II: a) computations by Runge-Kutta method for the system (45), (46) after spatial discretization, 
    b) computations by Crank-Nicolson method for the system (48), (49) after temporal and spatial discretization.}

\label{bld_rozwarcia}
\end{figure}

We compare the accuracy of the solution, obtained by the standard
Runge-Kutta MATLAB solver \textit{ode15s} for the dynamic system,
with that, obtained by the Crank-Nicolson method with iterations on
a time step for the discretized system (48), (49) and discretized
boundary conditions (34) and (35). In both cases, we used the same
second order approximations for the first and second spatial
derivatives. The number $N$ of nodal points uniformly spaced on the
interval $[0,1-\varepsilon ]$ and the regularization parameter $\varepsilon $%
, were also the same: $N=100$ and $\varepsilon =10^{-4}$. Time
stepping in
the second approach was taken similar to that generated by the solver \textit{%
ode15s} for solving the dynamic system of the first approach. The
time interval $[1,100]$ was also the same for both approaches. Thus
the conditions for the comparison were practically the same.
Fig.~\ref{bld_rozwarcia} illustrates the comparative accuracy of the
two approaches. It presents the relative error $\delta w$ of the
opening obtained by the first (Fig. \ref{bld_rozwarcia}a) and the
second (Fig. \ref{bld_rozwarcia}b) approaches for the benchmark
solution II. We see that the
both approaches provide high accuracy: the relative error does not exceed $%
1.8\cdot 10^{-6}$. Still, the Crank-Nicolson scheme provides
two-order less
error ($\delta w<10^{-8}$), except for the time close to the initial moment (%
$t_{0}=1$) and nodes closest to the crack tip ($\varsigma =1$).
(Surely, the error of the approach, using the Crank-Nicolson scheme,
for small time and near the tip, may be decreased to the level
$10^{-8}$, as well, by decreasing the first time steps and
increasing the number of iterations within a time step).

The Crank-Nicolson scheme served us also to compare the accuracy of
results, obtained by using (i) the SE (49), and (ii) the global mass
balance. In the normalized variables, the latter has the form:
\[
x_{\ast }(t)\int_{0}^{1}w(\varsigma ,t)d\varsigma =x_{\ast
}(t_{0})\int_{0}^{1}w(\varsigma ,t_{0})d\varsigma
+\int_{t_{0}}^{t}q_{0}(t)dt-\int_{t_{0}}^{t}x_{\ast
}(t)\int_{0}^{1}q_{l}(\varsigma ,t)d\varsigma dt.
\]

The calculations show that, when employing the global mass balance,
the maximal relative error $\delta w$ of the opening becomes
$1.4\cdot 10^{-3}$. It is three-order greater than the
error, obtained when employing the SE. This can be explained by an additional error induced by numerical integration over the interval [0,1].
This implies that using the
\textit{local} SE instead of the \emph{global} mass balance is beneficial
for the accuracy even in 1D problems.

Finally we analyze the dependence of the accuracy on the distribution
of the fluid velocity along the fracture. Table 1 presents the
maximal relative error \ of the opening $\delta w$ and the fracture
length $\delta x_{\ast }$ for the benchmark solutions I, II and III.
The data are obtained by using the first approach. The second line
of the table contains the maximal relative variation $\gamma _{v}$
of the fluid velocity. It can be seen that the variation notably
influences the accuracy. Even in the cases of the benchmark
solutions I and II, when the variation $\gamma _{v}$\ is of the same
order, there is significant (an order) difference in the accuracy.
For strongly non-uniform distribution, corresponding to the
benchmark solution III, the relative error is three orders greater
than that for the most uniform distribution, corresponding to the
benchmark solution II. This implies that numerical modeling of
fractures with notable variations of the fluid velocity requires
tests with growing density of the spatial mesh.

\vspace{3mm}

\begin{table}[h]
\begin{center}
\begin{tabular}{cccccc}
\toprule
$ $ & $ benchmark\, I $ & $ benchmark \,II $ & $ benchmark \,III$ \\[1mm]
\hline
\hline\\[-3mm]
$\gamma_{ v}$ & 0.06 & 0.02 &
0.4 \\
 \midrule
$\delta  w$ & $4.86 \cdot 10^{-5}$ & $1.54 \cdot 10^{-6}$ &
$5.04 \cdot 10^{-3}$\\
 \midrule
 $\delta x_*$ & $7.23 \cdot 10^{-5}$ & $1.97 \cdot 10^{-6}$ &
$6.84 \cdot 10^{-3}$\\
\bottomrule
\end{tabular}
\end {center}
\caption{Accuracy of the crack opening $\delta  w$ and crack length $\delta x_*$ for various benchmark solutions.} \label{table1}
\end{table}

\vspace{-3mm}

\section{Sensitivity analysis}

For zero leak-off, the only parameter, defining the solution, is the
influx at the inlet $q_{0}$. Consider its perturbed value
$q_{0}+\Delta q_{0}$, so that the relative change of the influx is
$\delta q_0=\Delta q_{0}/q_{0}$. We are interested in finding relative changes
of the opening $\delta w=\Delta w/w$, the pressure $\delta p=\Delta p/p$, the fracture
length $\delta x_*=\Delta x_{\ast }/x_{\ast }$, the particle velocity $\delta v=\Delta
v/v$ and the front speed $\delta V=\Delta V_{\ast }/V_{\ast }$. From the
analytical solution (38), (39), we easily obtain:
\[
{\delta w}=\delta p=(1+\delta \xi _{\ast
})^{2/3}-1, \quad \delta v=\delta V_{\ast }=%
\delta x_{\ast }=\delta \xi _{\ast }.
\]

In Sec. 2, it was stated that $\xi _{\ast }$ is proportional to $%
q_0^{3/5}$. Hence, $1+\delta \xi _{\ast }=(1+
\delta q_{0})^{3/5}$, and finally the relative changes are:
\[
\delta w=\delta p=(1+\delta q_{0})^{2/5}-1, \quad \delta v=\delta V_{\ast }=
\delta x_{\ast }=(1+\delta
q_{0})^{3/5}-1.
\]

For a small relative change of the influx, to the accuracy of terms
of order $O\left(\delta q_{0}^3\right)$, we
have:
\begin{equation}
\delta w=\delta p=0.4\delta q_{0}-0.12\delta q_{0}^2,\quad \delta v=\delta V_{\ast }=\delta x_{\ast }=0.6\delta q_{0}
-0.12\delta q_{0}^2.
\end{equation}

This implies that the relative changes equal approximately to
one-half of the relative change of the influx. Note that the
disturbances of the solution change its sign when $\delta q_{0}$
changes the sign.

\begin{figure}[h!]

    \hspace{-3mm}\includegraphics [scale=0.38]{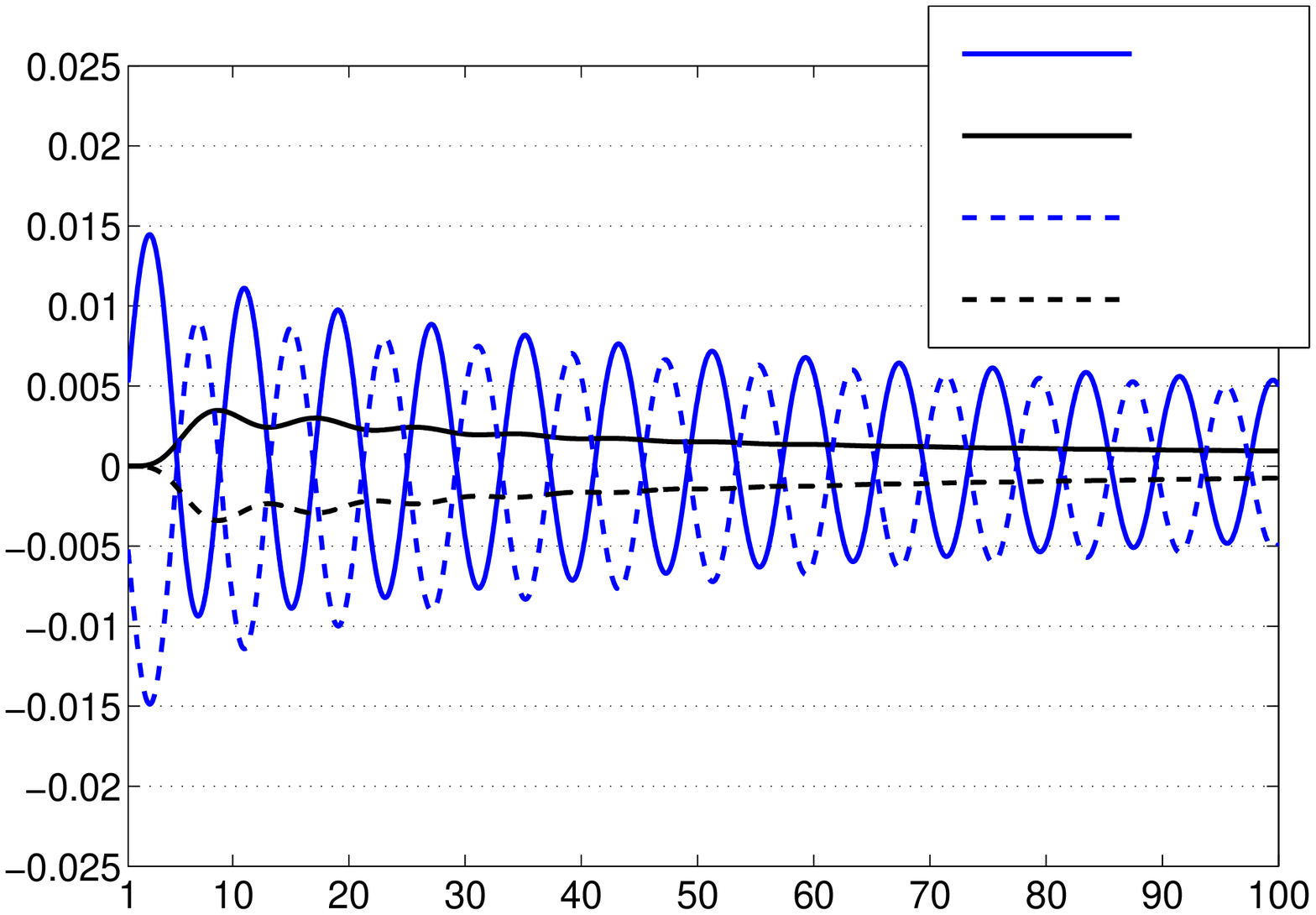}
    \put(-115,0){$t$}
    \put(-39,145){$\delta w_0$}
    \put(-39,133){$\delta x_*$}
    \put(-39,120){$\delta w_0$}
    \put(-39,107){$\delta x_*$}
    \put(-235,140){a)}
    \hspace{-2mm}\includegraphics [scale=0.38]{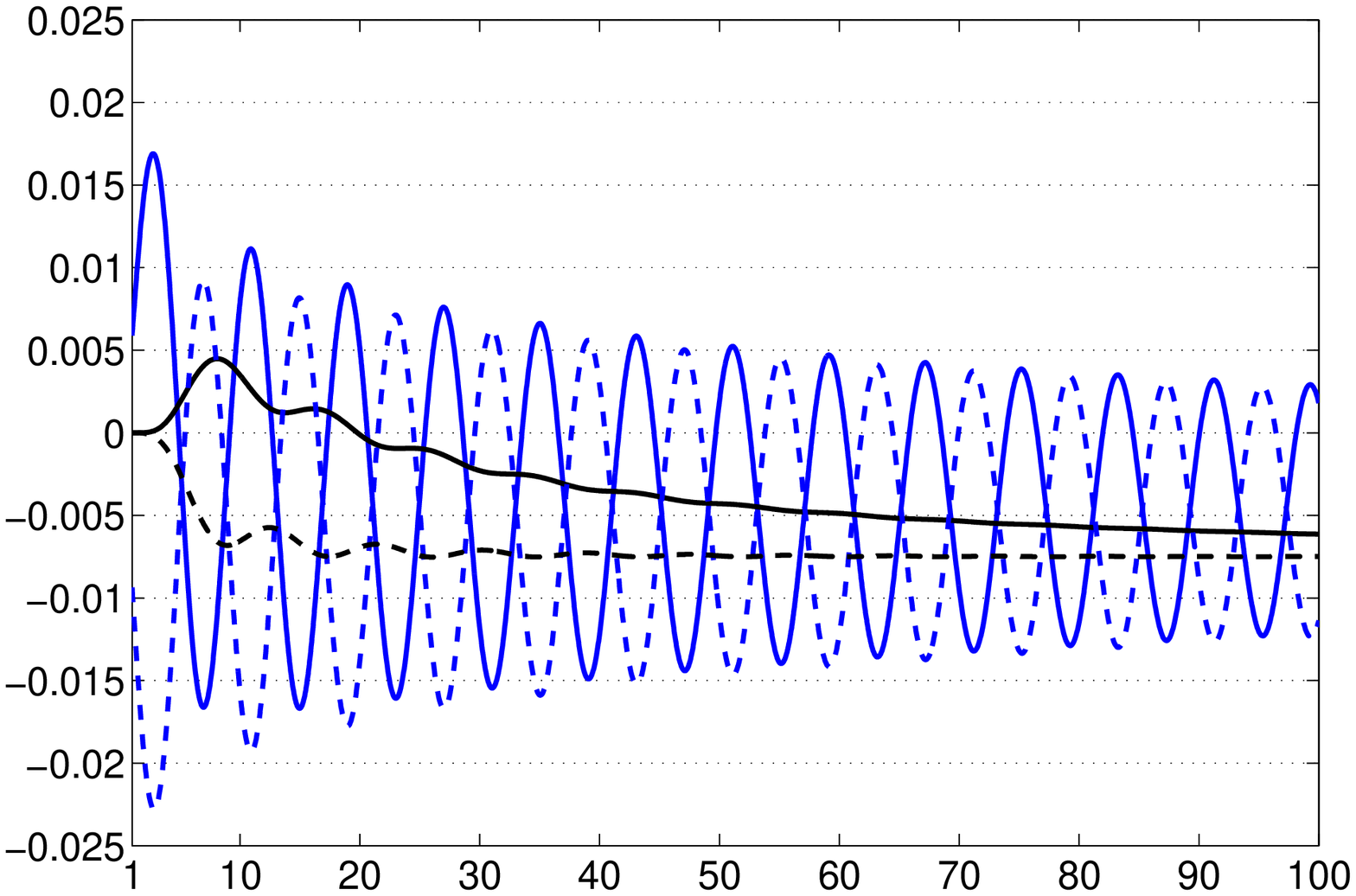}
    \put(-115,0){$t$}
   \put(-235,140){b)}

    \caption{Relative fluctuations of the crack opening $\delta w_0$ and the crack length $\delta x_*$, 
    caused by the periodic disturbance of the influx: a) benchmark solution II, b) benchmark solution III.}

\label{sens}
\end{figure}

In cases, when the disturbance $\delta q_{0}$\ oscillates in time
with the amplitude $A$ and the angular frequency $\omega _{q}$\ as
$\delta q_{0}=A\sin (\omega _{q}(t-t_{0}))$, we may numerically
evaluate its influence on the solution by employing any of the
discussed approaches. The results, obtained when solving the
discretized dynamic system (45), (46) by using standard MATLAB
solver \textit{ode15s} are presented in Fig. \ref{sens}.

\vspace{2mm}

The graphs
are plotted for the relative amplitude of perturbation $A/q_{0}=\pm
0.1$ and $\omega_q=\pi/4$. They present the relative fluctuations of the
crack opening, $\delta w_0=\Delta w(0,t)/w(0,t)$, at the inlet
point ($\varsigma=0$) and relative fracture length $\delta x_{\ast}$
in time for the benchmark solution II (Fig. \ref{sens}a) and III (Fig. \ref{sens}b).
Solid lines refer to the case when
$A/q_{0}=+ 0.1$,  dashed lines to $A/q_{0}=- 0.1$.
It can be seen that for the periodic perturbations the crack opening $\delta w_0$ is an order less than the amplitude $A$ of the perturbation $\delta q_0$ itself,
while the position of the crack, $\delta x_*$, is two orders less.
It is worth noting that the fluctuations depend on the sign of the amplitude $A$, and the change
of the fracture length tends to a constant value with growing time.
It can be seen that, when the amplitude $A$ changes its sign, the limit of the fluctuation of the fracture length for small leak-off (benchmark II)
also changes its sign, while for the large leak-off (benchmark III) the sign is the same.

\section{Conclusions}

The results presented show that the modified formulation of the
problem 
\cite {Linkov_1} -- \cite{Linkov_4}, 
employing the suggested variables, speed equation
and $\varepsilon$-regularization, extends the opportunities for better numerical
modeling of the hydraulic fracture propagation. Specifically, the
improvements may include:

(i) drastic decrease of stiffness of ODE, obtained after spatial
discretization, by employing iterations in the particle velocity;

(ii) avoiding iterations in non-linear terms and using highly efficient standard solvers by including the SE into the system of ODE as suggested in Sec. 3;


(iii) increasing the accuracy of time-stepping schemes of
Crank-Nicolson type by using the SE instead of the commonly used
global mass balance.

The advantages of the modified formulation have been illustrated by
considering the simplest 1D problem by Nordgren. Meanwhile, the
suggested formulation and approaches, being general, they are
applicable to 2D fractures. They are of special interest for
modeling the area behind the fluid front, where gradients of the
pressure and opening are high.

\subsubsection*{Acknowledgements}

This research has been supported by FP7 Marie Curie IAPP project
(PIAP-GA-2009-251475). MW and AM are grateful respectively to the Institute of
Mathematics and Physics of Aberystwyth University and EUROTECH for the facilities
and hospitality.


\bigskip

\begin{thebibliography}{99}
\bibitem{AdachiDetour-2002}  Adachi, J., Detournay, E. 2002. Self-similar
solution of a plane-strain fracture driven by a power-law fluid. \textit{%
Int. J. Numer. Anal. Methods Geomech}. 26, 579--604.

\bibitem{Adachi-et-Al-2007}  Adachi, J., Siebrits E., Peirce A., Desroches
J. 2007. Computer Simulation of Hydraulic Fractures.
\textit{International Journal of Rock Mechanics and Mining
Sciences}, 44, 739-757.

\bibitem{BungerDetourGarag-2005}  Bunger, A.P., Detournay, E. \& Garagash,
D.I. 2005. Toughness-dominated hydraulic fracture with leak-off.
Int. J. Fracture, 134, 175-190.

\bibitem{Carter}  Carter, E. 1957. Optimum fluid characteristics for
fracture extension. In: Howard, G., Fast, C. (eds.) \textit{Drilling
and Production Practices}, pp. 261--270. American Petroleum
Institute.

\bibitem{Desroches-et-al-1994}  Desroches, J., Detournay, E., Lenoach, B.,
Papanastasiou, P., Pearson, J., Thiercelin, M., Cheng, A.-D. 1994.
The crack tip region in hydraulic fracturing. \textit{Proc. Roy.
Soc. Lond. Ser. A} 447, 39--48 (1994)

\bibitem{Detournay-2004}  Detournay, E. 2004. Propagation regimes of
fluid-driven fractures in impermeable rocks. \textit{Int. J.
Geomech}. 4(1), 1--11.

\bibitem{Economides2000}  Economides, M., Nolte, K. (eds.). 2000. Reservoir
Stimulation. 3rd edn. Wiley, Chichester, UK.

\bibitem{Garagash2006}  Garagash, D. I. 2006. Propagation of a plane-strain
hydraulic fracture with a fluid lag: Early time solution. Int. J.
Solids Struct. 43, 5811-5835.

\bibitem{GaragashDetour-2000}  Garagash, D., Detournay, E. 2000. The tip
region of a fluid-driven fracture in an elastic medium. ASME J.
Appl. Mech. 67(1), 183--192 (2000).

\bibitem{GaragDetourAdachi-2011}  Garagash, D.I., Detournay, E. \& Adachi,
J. I. 2011. Multiscale tip asymptotics in hydraulic fracture with
leak-off. J. Fluid Mech., 669, 260-297.

\bibitem{Geertsma and de Klerk (1969)}  Geertsma, J. \& de Klerk, F. 1969. A
rapid method of predicting width and extent of hydraulically induced
fractures. J. Pet. Tech., 21, 1571-1581.

\bibitem{Hadamard-1902}  Hadamard, J. 1902. Sur les problemes aux derivees
partielles et leur signification physique. Princeton University
Bulletin, 49-52.

\bibitem{Higham-1986}  Higham, N. 1986. Efficient algorithms for computing
the condition number of a tridiagonal matrix. The University of
Manchester, MIMS EPrint 2007.8.

\bibitem{Howard and Fast (1969)}  Howard, G.C. \& Fast, C.R. 1970. Hydraulic
fracturing. Monograph Series Soc. Petrol. Eng., Dallas.

\bibitem{Hu-Garagash-2010}  Hu, J. \& Garagash, D.I. (2010). Plane strain
propagation of a fluid-driven crack in a permeable rock with
fracture toughness. ASCE J. Eng. Mech., 136, 1152-1166.

\bibitem{Jamamoto et al. (2004)}  Jamamoto, K. \& Shimamoto, T. \& Sukemura,
S. 2004. Multi fracture propagation model for a three-dimensional
hydraulic fracture simulator. Int. J. Geomech. ASCE, 1, 46-57.

\bibitem{KovalyshenDet2009}  Kovalyshen, Y. and Detournay, E. 2009. A
reexamination of the classical PKN model of hydraulic fracture, \textit{%
Transp Porous Med}, 81, 317-339.

\bibitem{Khristianovich and Zheltov (1955)}  Khristianovich, S.A. \&
Zheltov, V.P. 1955. Formation of vertical fractures by means of
highly viscous liquid. In: Proc. 4-th World Petroleum Congress,
Rome, 579-586.

\bibitem{Linkov_1}  Linkov, A.M. 2011. Speed equation and its application
for solving ill-posed problems of hydraulic fracturing. \textit{ISSM
1028-3358, Doklady Physics}, Vol.56, No.8, pp.436-438. Pleiades
Publishing, Ltd. 2011.

\bibitem{Linkov_2}  Linkov, A.M. Use of a speed equation for numerical
simulation of hydraulic fractures. Available at:
http://arxiv.org/abs/1108.6146. Date: Wed, 31 Aug 2011 07:47:52 GMT
(726kb). Cite as: \textit{arXiv}: 1108.6146v1 [physics.flu-dyn].

\bibitem{Linkov_3}  Linkov, A.M. 2011. On numerical simulation of hydraulic
fracturing. \textit{Proc. XXXVIII Summer School-Conference,
\textquotedblleft Advanced Problems in
Mechanics-2011\textquotedblright }, Repino, St. Petersburg, July
1-5, 2011, 291-296.

\bibitem{Linkov_4}  Linkov, A.M. 2012. On efficient simulation of hydfraulic
fracturing in terms of particle velocity. \textit{Int. J. Eng.
Sci}., 52, 77-88.

\bibitem{Mitchell2007}  Mitchell, S.L., Kuske, R., Peirce, A. 2007. An
asymptotic framework for finite hydraulic fractures including
leakoff. SIAM J. Appl. Math. 67(2), 364--386.

\bibitem{Nolte (1988)}  Nolte, K.G. (1988). Fracture design based on
pressure analysis. Soc. Pet. Eng. J., Paper SPE 10911, February,
1-20.

\bibitem{Nordgren}  Nordgren R.P. 1972. Propagation of a Vertical Hydraulic
Fracture. \textit{J. Pet. Tech}, 253, 306-314.

\bibitem{Perkins}  Perkins, T., Kern, L. 1961. Widths of hydraulic
fractures. J. Pet. Tech. Trans. AIME 222, 937--949.

\bibitem{Savitski Detour-2002}  Savitski, A. \& Detournay, E. 2002.
Propagation of a fluid driven penny-shaped fracture in an
impermeable rock: asymptotic solutions. Int. J. Solids Struct. 39,
6311-6337.

\bibitem{Sethian-1999}  Sethian, J.A. 1999. Level set methods and fast
marching methods. Cambridge, Cambridge University Press.

\bibitem{Spence&Sharp}  Spence, D.A., Sharp P. W. 1985. Self-similar
solutions for elastohydrodynamic cavity flow. \textit{Proc. Roy.
Soc. London, Series A}, 400, 289-313.
\end{thebibliography}
\end{document}